# Stable reconstruction of the (110) surface and its role in pseudocapacitance of rutile-like $RuO_2$

**Hayk A. Zakaryan [1,+] Alexander G. Kvashnin, [2,3,*,+] Artem R. Oganov [2,3,4,5]**

[1] Yerevan State University, 1 Alex Manoogian St., 0025, Yerevan, Armenia
[2] Skolkovo Institute of Science and Technology, Skolkovo Innovation Center 143026, 3 Nobel Street, Moscow, Russian Federation
[3] Moscow Institute of Physics and Technology, 141700, 9 Institutsky lane, Dolgoprudny, Russian Federation
[4] Department of Geosciences and Center for Materials by Design, Institute for Advanced Computational Science, State University of New York, Stony Brook, NY 11794-2100
[5] International Center for Materials Discovery, Northwestern Polytechnical University, Xi'an, 710072, China
*A.Kvashnin@skoltech.ru
[+]these authors contributed equally to this work

**ABSTRACT**

Surfaces of rutile-like $RuO_2$, especially the most stable (110) surface, are important for catalysis, sensing and charge storage applications. Structure, chemical composition, and properties of the surface depend on external conditions. Using the evolutionary prediction method USPEX, we found stable reconstructions of the (110) surface. Two stable reconstructions, $RuO_4$–(2×1) and $RuO_2$–(1×1), were found, and the surface phase diagram was determined. The new $RuO_4$–(2×1) reconstruction is stable in a wide range of environmental conditions, its simulated STM image perfectly matches experimental data, is more thermodynamically stable than previously proposed reconstructions, and explains well pseudocapacitance of $RuO_2$ cathodes.

**Introduction**

In the era of nanotechnology, steady miniaturization of electronic devices to nanometer scale takes place, with quantum and surface effects playing a major role for properties and stability. Surface science becomes crucial for future. One of the most studied materials for catalysis, sensing and energy applications is the rutile-type $RuO_2$. [1] Many researchers studied catalytic properties of $RuO_2$ to enhance its catalytic efficiency for the oxidation of CO, NO and other molecules, which are important in industry. [2–4] In sensing devices, ruthenium is often used as a dopant for rutile-type $SnO_2$. Ruthenium oxide is used in many applications as a thin layer to enhance sensitivity and selectivity of devices. [5,6] It was also used as a cathode material for supercapacitors, displaying constant capacitance over the wide range of electric potential. [7]

All these applications rely on unique properties of ruthenium dioxide. Under normal conditions $RuO_2$ has tetragonal rutile-type structure, with $P4_2/mnm$ space group with two ruthenium and four oxygen atoms in the unit cell. [8,9] Under high pressures $RuO_2$ transforms to a $CaCl_2$-type phase at 6 GPa [10] and to pyrite structure at 82 GPa. [11]

There are also several known ruthenium oxides: $RuO_4$, RuO and $RuO_3$. [12] RuO exists in a gas phase at temperatures above 1900 K. [12,13] Ruthenium trioxide ($RuO_3$) exists in a gaseous form in the temperature range from 1300 to 2000 K, while the solid state of $RuO_3$ forms only on substrates, i.e. on quartz surface at 400 K. [12] Ruthenium tetroxide can be in a gas, liquid or solid states. Below 1300 K the gaseous ruthenium tetroxide ($RuO_4$) is formed, [12] which condenses at temperatures below 300 K. [12,14]

A number of theoretical and experimental works were devoted to detailed investigation of different surfaces of $RuO_2$. [15–17] It was found that at ambient conditions the most stable $RuO_2$ surface has (110) crystallographic orientation. [18] However, the atomic structure and even the composition of the surface can be changed under different environmental conditions (temperature and partial pressure of oxygen). [18] Several theoretical predictions of possibly stable terminations of (110)-$RuO_2$ surface were made by Reuter et al. [19,20]

Scanning tunneling microscopy (STM) is often used to study surfaces of materials.[15] However, it shows only the top layers of the materials and in a case of $RuO_2$ only oxygen can be distinguished. [15] Thus, the actual structure of the surface becomes largely hidden from the eye of the experimentalist. Due to the fact that (110)-$RuO_2$ surface is very sensitive to environmental conditions, the atomic structure and stoichiometry of the (110) surface may change drastically. [15]

$RuO_2$ is the most widely used material in pseudocapacitors, novel energy storage devices, which are in great demand for different applications. The pseudocapacitive behavior of $RuO_2$ was first studied and explained by Trasatti and Buzzanca. [21]In their paper, it was proposed that the main mechanism of charge storage can be explained by the following redox reaction:

$$RuO_x(OH)_y + \delta H^+ + \delta e^- \leftrightarrow RuO_{x-\delta}(OH)_{y+\delta}. \quad (1)$$

Supercapacitive behavior occurs due to proton-electron double insertion. Thus each adsorbed or intercalated hydrogen atom (proton) will induce pseudocapacitance in the cathode material. Despite intense research devoted to the study of hydrogen intercalation in the cathode materials, [22–29] the atomic-scale processes are still not clearly understood. One of the main problems is the influence of proton adsorption, because it is difficult to distinguish surface pseudocapacitance (charge stored due to protons intercalation into the material) from double layer capacitance (charge stored due to electrostatic potential between electrode surface and electrolyte). The energetics of proton intercalation, atomic structure and stability of hydrogenated surface are still uncertain. For all of these problems, an investigation of possible surface reconstructions is essential.

It should be noted that none of prior theoretical predictions used global optimization techniques to find the most stable reconstructions of the (110)-$RuO_2$ surface. Using evolutionary structure prediction algorithm USPEX [30–33] and density functional theory we discovered new reconstructions of the (110)-$RuO_2$ surface. This allows clearer explanations and deeper understanding of the processes occurring on surfaces. The formation conditions of studied reconstructions were estimated by the calculations of the surface energy as a function of oxygen chemical potential. Obtained phase diagram gives stability fields of different reconstructions in terms of various environmental conditions (oxygen partial pressure and temperature). Calculated voltage for adsorption of hydrogen on the new (110)-$RuO_2$ surface reconstructions will answer the question “how the surface redox reaction contributes to pseudocapacitance of $RuO_2$ electrode?”.

## Results

We searched for stable reconstructions of (110)-$RuO_2$ surface using variable-composition evolutionary algorithm USPEX adapted for surfaces. [33] We predicted several reconstructions, shown in Figure 1. It is important to note that all these reconstructions have the same substrate, and the surface reconstruction takes place on top of the substrate, in the thickness region 3-5 Å.

We found 2 stable and 2 metastable reconstructions, which are closest to convex hull (see Figure 2b). Different reconstructions of (110)-$RuO_2$were denoted as $RuO_4$–(2×1) (Figure 1a), $RuO_2$–(1×1) (Figure 1b), $Ru_4O_9$–(1×1) (Figure 1c) and $Ru_8O_{17}$–(1×2) (Figure 1d). The nomenclature of the predicted reconstructions reflects the stoichiometry of reconstructed surface regions. The stoichiometry of the surface region equals the difference between stoichiometry of the entire system minus stoichiometry of the substrate. Number in the brackets is the number of surface cells in the reconstructed cell. The total number of the atoms in considered structures can be found in Table 1.

Three of predicted reconstructions have already been known from previous theoretical studies: $RuO_2$–(1×1), $Ru_4O_9$–(1×1) and $Ru_8O_{17}$–(1×2). [20,34] Reconstruction $RuO_4$–(2×1) is newly predicted. It is interesting to note, that among all predicted reconstructions we found one ($Ru_4O_5$–(1×1)), which contains a RuO monolayer on top of the $RuO_2$ substrate.

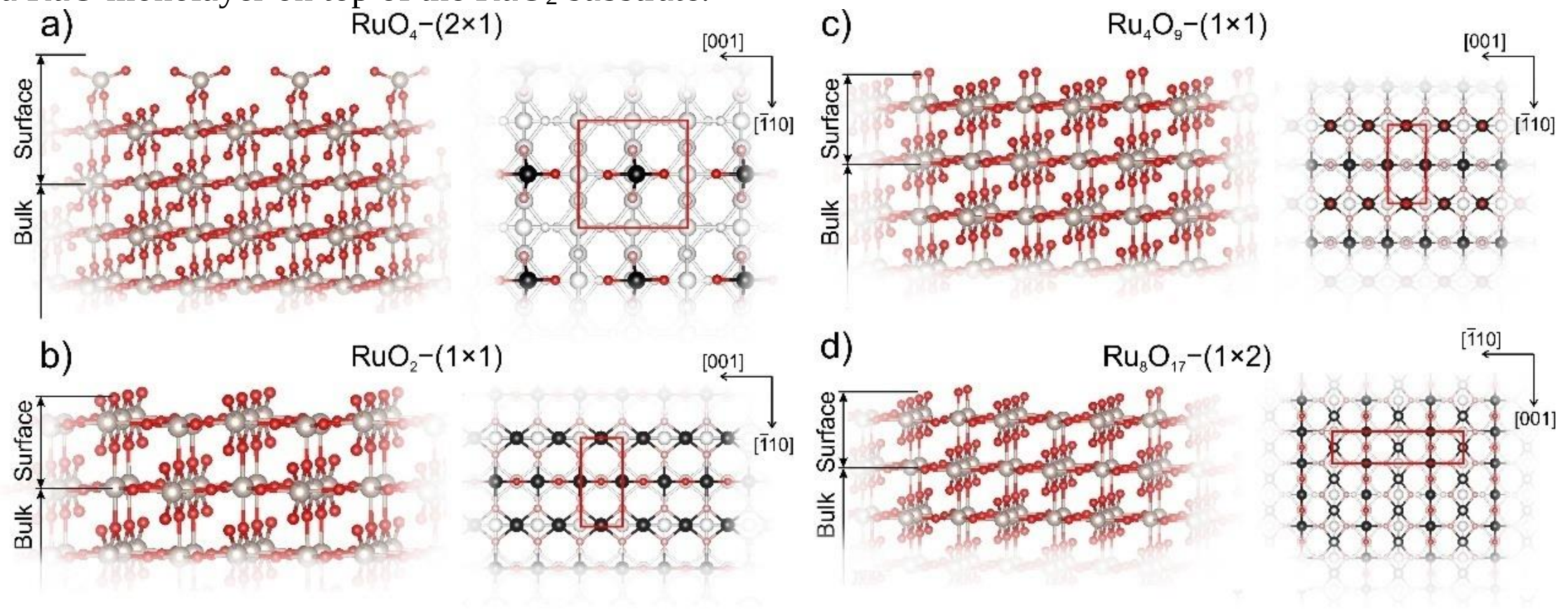

**Figure 1.** Predicted new reconstructions of (110)-$RuO_2$ surface: stable a) $RuO_4$–(2×1), b) $RuO_2$–(1×1), and closest to convex hull metastable c) $Ru_4O_9$–(1×1) and d) $Ru_8O_{17}$–(1×2). In the top views the Ru atoms of the top layer are black, oxygen atoms of the upper layer are red, oxygen atoms following the top layer are light red.

Let us now move to investigation of stability of predicted surface reconstructions. Using equation (5) we calculated the surface energy of all predicted surface reconstructions as a function of oxygen chemical potential, shown in Figure 2a. The structure with the lowest surface energy in a given range of chemical potentials is deemed stable at those chemical potentials. The range of oxygen chemical potential from -6.45 to -4.95 eV is experimentally achievable (see eq. (6)). For the convenience of readers, we placed all the equations in the Methods section in the end of the paper. Chemical potential lower than -6.45 eV will lead to desorption of oxygen and pure ruthenium will precipitate. Values $\mu_O > -4.95$ eV indicate the formation of oxygen molecules ($O_2$) on the surface. As one can see from Figure 2a, there are two stable surface reconstructions: $RuO_2$–(1×1) and $RuO_4$–(2×1) (dashed blue and bold red lines in Figure 2a). $RuO_2$–(1×1) is stable in the range of oxygen chemical potentials from -6.45 to -5.84 eV and has bulk stoichiometry and bulk-like termination (Ru:O=1:2). The other stable structure is $RuO_4$–(2×1), which has the lowest surface energy ($G_s$) in the range of $\mu_O$ from -5.84 to -4.95 eV (red line in Figure 2a). This reconstruction has one four-coordinate Ru atom and 4 oxygen atoms, two of which are two-coordinate and the other two are one-coordinate (Figure 1a). According to Ref. [20], another reconstruction called "Cusp" (in our study it is $Ru_4O_9$–(1×1) due to another nomenclature) should be stable in the same range as our $RuO_4$–(2×1). $Ru_4O_9$–(1×1) has bulk-like termination with one additional oxygen atom located on top of 5-coordinate Ru atom (see red atom in the top view of Figure 1c). We found that $Ru_4O_9$–(1×1) reconstruction has surface energy higher than $RuO_4$–(2×1) by 0.1 eV (see dotted line in Figure 2a) and therefore is metastable. Here and below all energy values are taken per unit cell. It is important that $RuO_4$–(2×1) and $Ru_4O_9$–(1×1) reconstructions have the same $\Delta N$, which leads to the same slopes of $G_s(\mu_O)$ functions (they are parallel). Values of $\Delta N$ and $\Delta E$ for all predicted surface reconstructions calculated by using eq. (7) are presented in Table 1. Additional calculations with a doubled substrate thickness along *c*-axis gave the same result, i.e. stability of $RuO_4$–(2×1) versus $Ru_4O_9$–(1×1).

**Table 1.** Predicted surface reconstructions. Number of ruthenium and oxygen atoms ($N_{Ru}$, $N_O$), number of multiplications of the unit cell ($N_{cell}$), total energy per cell from DFT calculations ($E_{total}$), $\Delta N$ and $\Delta E$ values.

| Structure | $N_{Ru}$ | $N_O$ | $N_{cell}$ | $E_{total}$, eV | $\Delta N$ | $\Delta E$, eV |
|---|---|---|---|---|---|---|
| $RuO_4$-(2×1) | 17 | 36 | 2 | -383.24 | 1 | -3.77 |
| $Ru_4O_9$-(1×1) "cusp" | 8 | 17 | 1 | -180.39 | 1 | -3.59 |
| $Ru_8O_{17}$-(1×2) | 16 | 33 | 2 | -355.17 | 0.5 | -0.78 |
| $RuO_2$-(1×1) | 8 | 16 | 1 | -174.73 | 0 | 2.07 |

The convex hull diagram for all considered structures is shown in Figure 2b, where each point represents one structure. Solid points represent thermodynamically stable reconstructions, which form the convex hull. Metastable reconstructions are open circles and are located above the convex hull. Here only two reconstructions are found to be stable, namely $RuO_2$–(1×1) and $RuO_4$–(2×1). $Ru_8O_{17}$–(1×1) and $Ru_4O_9$–(1×1) are located very close to the convex hull line just by 0.06 and 0.1 eV above it, respectively.

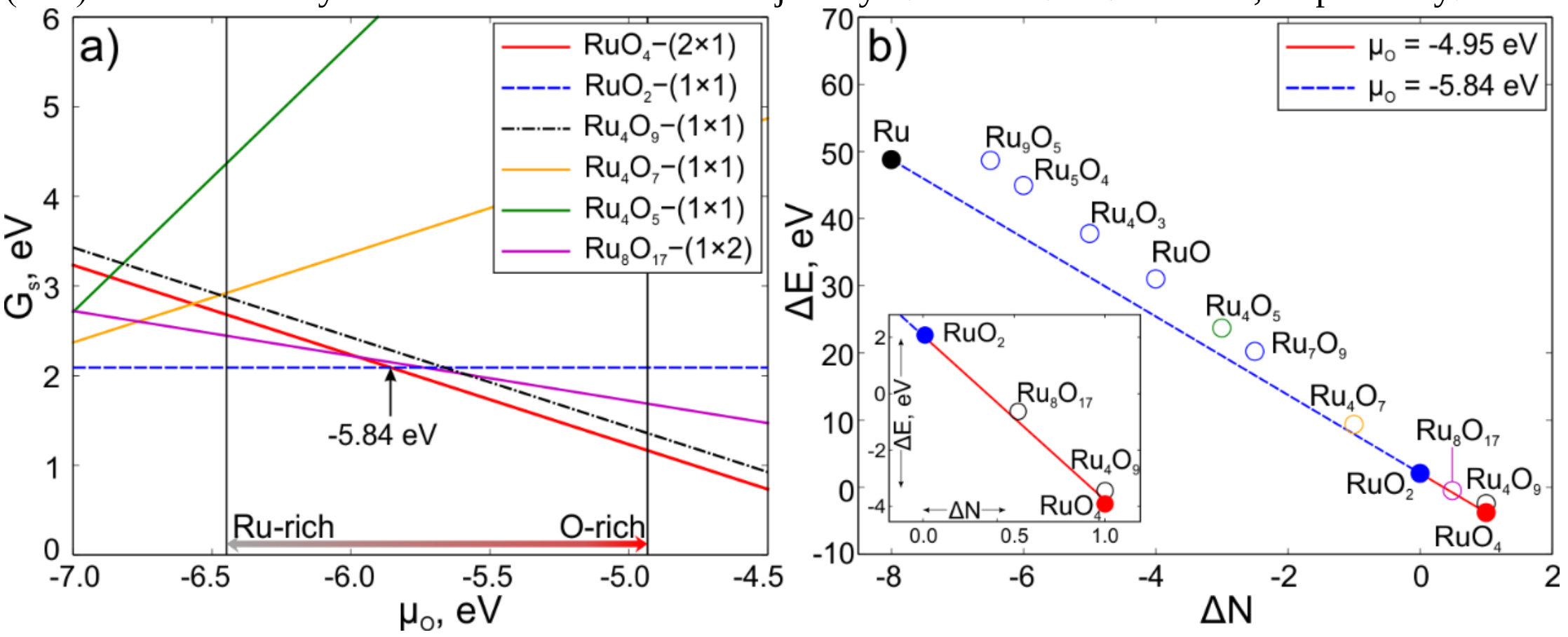

**Figure 2.** a) Surface energy per unit cell as a function of oxygen chemical potential ($\mu_O$). b) Convex hull of (110)-$RuO_2$ reconstructions. Color of points corresponds to the color of lines in a). The inset zooms in on the region of $\Delta N$ from -0.5 to 1.5.

The metastable reconstruction $Ru_8O_{17}$–(1×2) is geometrically similar to $Ru_4O_9$–(1×1), but with doubled cell in the $[\bar{1}10]$ direction, and one oxygen removed from a site above Ru (see Figure 1d). This reconstruction has energy 0.06 eV above the convex hull (see Figure 2b). $Ru_4O_7$–(1×1) is unstable, as was already shown in previous studies. [20]

To discriminate between structural models, we use the results of Scanning Transmission Microscopy (STM). We simulated the STM images of $RuO_4$–(2×1), $RuO_2$–(1×1) and $Ru_4O_9$–(1x1) reconstructions as the most stable ones. The comparison between them and experimental STM image of $RuO_2$(110) surface was made. In Figure 3a, simulated STM image of $RuO_4$–(2×1) is presented, where bright dots are one-coordinate oxygen atoms. The distance between the atoms along the [001] direction (yellow arrow in Fig. 3a) is 3.2 Å, while the distance in the perpendicular direction (between the rows of atoms) is 6.26 Å. STM image of $Ru_4O_9$–(1×1) reconstruction is shown in Figure 3b, where the distance along the [001] direction is 3.2 Å, in the perpendicular direction the distance equals to 6.4 Å. The simulated STM image of $RuO_2$–(1×1) is in Figure 3c, where distances along [001] direction and perpendicular to it are 3.2 and 6.4 Å, respectively. These images agree well with experimental data, where the corresponding distances are 3.12 and 6.38 Å, respectively [35,36] (see Figure 3d). One must admit that all three models generate STM images consistent with experiment, which makes them difficult to distinguish from each other in experiments. While equally consistent with experimental STM images, our $RuO_4$–(2×1) reconstruction is lower in energy and therefore is preferable.

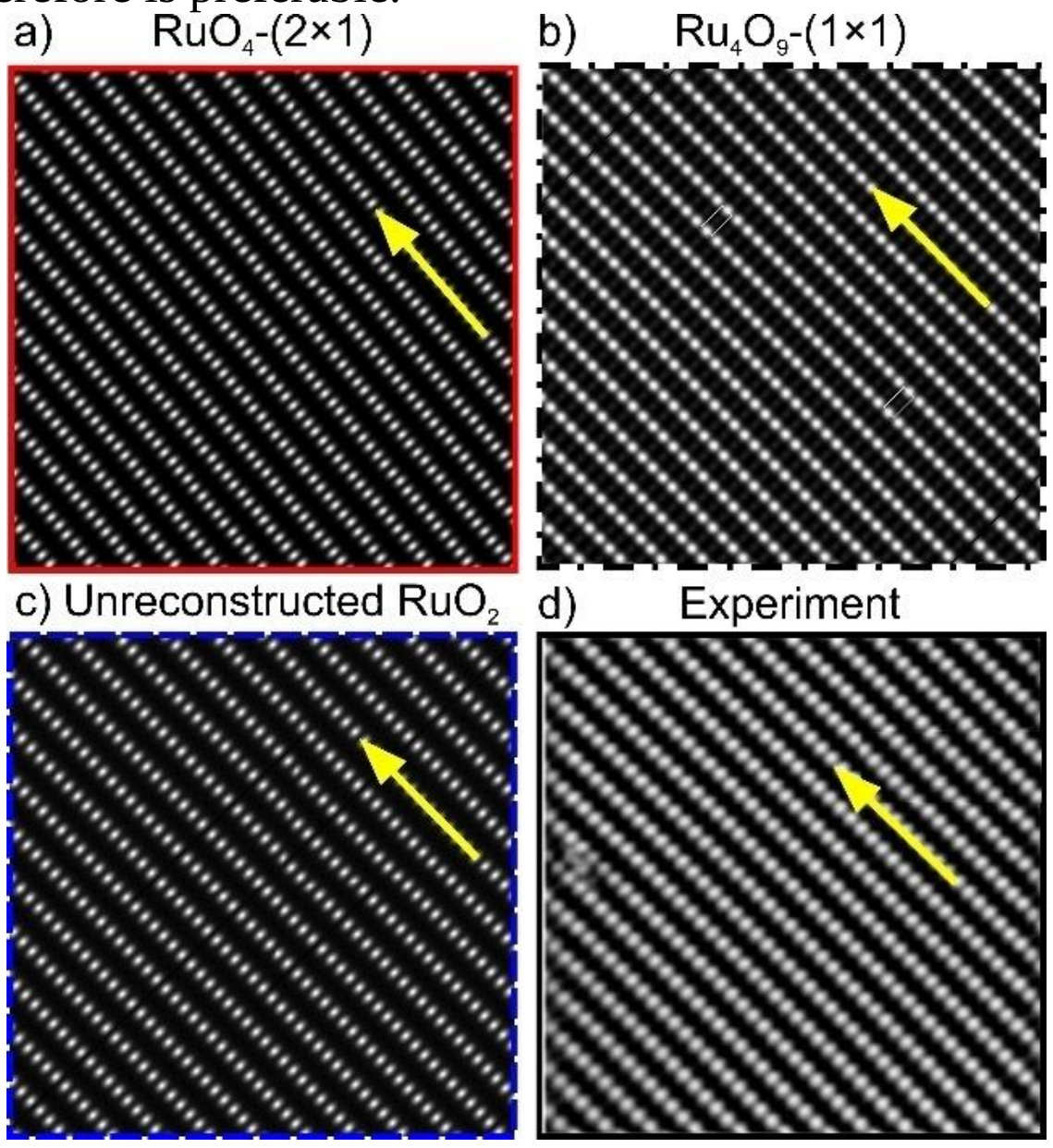

**Figure 3.** a) Simulated STM images of $RuO_4$–(2×1) and b) $Ru_4O_9$–(1×1) reconstructions; c) STM image of stoichiometric $RuO_2$ (110) surface; d) experimental STM image of $RuO_2$ surface from Ref. [36]. The [001] direction is highlighted by yellow arrows.

To determine stability fields of each surface reconstruction, we calculated the pressure-temperature phase diagram, shown in Figure 4. Such phase diagram shows environmental conditions (partial pressure and temperature), suitable for the formation of new reconstructions. Both partial pressure of oxygen and temperature both enter the expression for the chemical potential:

$$\mu_O = \frac{1}{2}\left[E_{O_2} + \Delta H_{O_2}(T,P_0) - TS_{O_2}(T,P_0) + k_B T ln\left(\frac{P}{P_0}\right)\right] = \frac{1}{2}E_{O_2} + \Delta\mu_O(T,P), \qquad (2)$$

where $E_{O_2}$ is the energy of the $O_2$ molecule (computed from first principles), $\Delta H_{O_2}(T,P_0)$, $TS_{O_2}(T,P_0)$ are thermal parts of the Gibbs free energy of the gas of oxygen molecules as a function of temperature and pressure, and it was taken from thermodynamic database. [37]

One can see from the calculated phase diagram (Figure 4) that $RuO_4$–(2×1) reconstruction is stable at higher values of oxygen partial pressure and lower temperatures than $RuO_2$–(1×1). The phase boundary (red curve in Figure 4) was plotted using equation (2) with the value of chemical potential of oxygen equal to $-5.84$ eV. Increasing temperature to 800 K at the pressure of $10^{-8}$ bar (dashed horizontal line) will lead to the formation of $RuO_2$–(1×1) reconstruction (see Figure 4).This fact perfectly agrees with experimental results, where $RuO_2$–(1×1) reconstruction forms at $log(p/p_0) = -8$ and $T \geq 600\ K$. [35,38] Further increase of temperature up to 1050 K leads to desorption of oxygen (see blue line in Figure 4). Blue line was plotted using $\mu_O = -6.45$ eV in equation (2). Oxygen chemical potential equals to $-6.45$ eV delineates the region where the deposition of pure ruthenium is favored. Note that at ambient conditions our $RuO_4$-(2×1) reconstruction is the one which is stable.

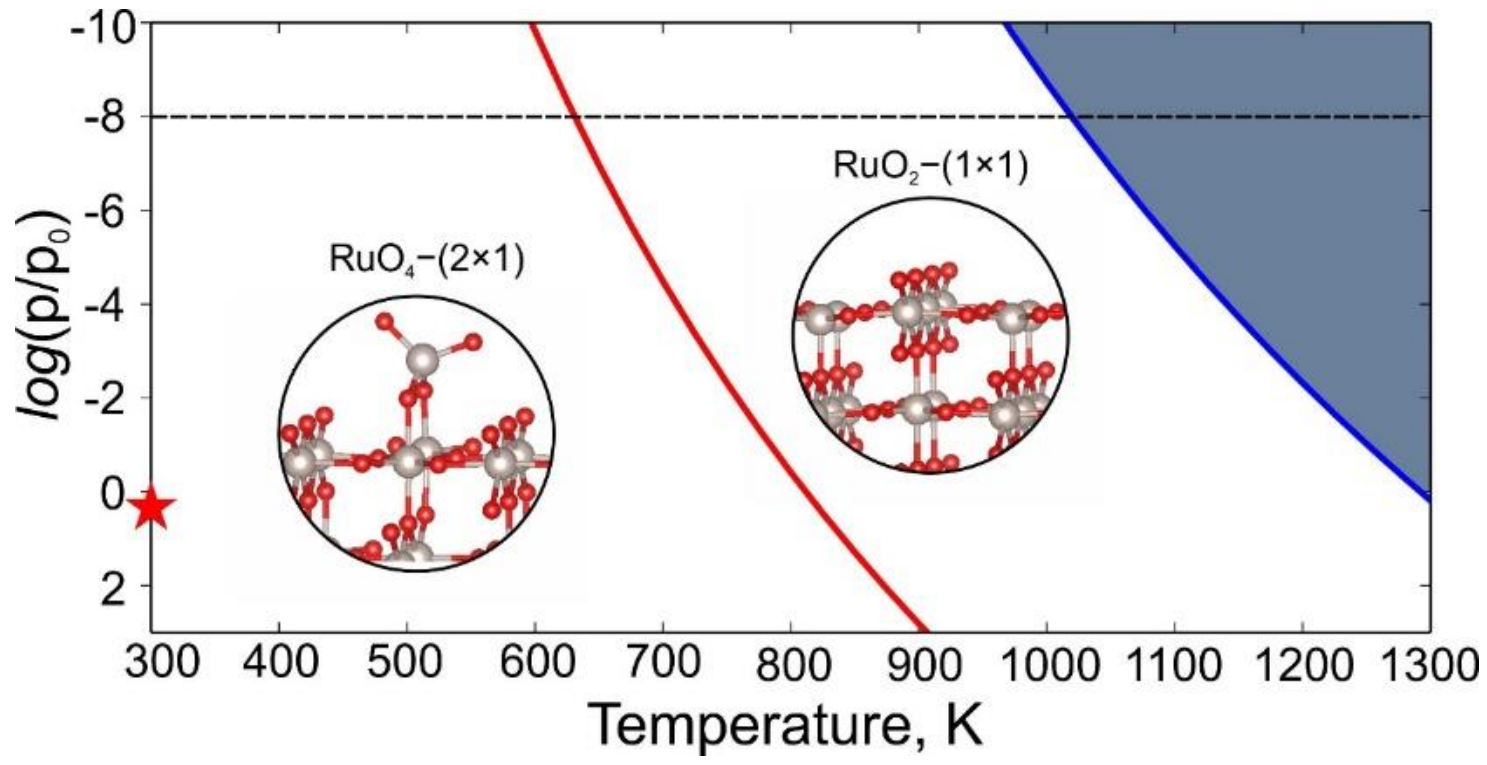

**Figure 4.** Surface phase diagram of (110)-$RuO_2$. The dark region corresponds to the deposition of Ru metal. The *x*-axis is temperature; *y*-axis is oxygen partial pressure. The star denotes ambient conditions ($p \approx 0.21$ atm, $T = 273$ K).

Thermal stability of the newly predicted $RuO_4$–(2×1) reconstruction was studied by means of molecular dynamics (MD) simulations. MD simulations were carried out at temperatures of 500 and 1200 K using the Nosé–Hoover thermostat [39,40] with a time step of 1 fs for a total simulation period of 5 ps. During the simulations the atomic structure of $RuO_4$–(2×1) surface reconstruction remains intact: essentially, only dynamical bending of the O-Ru-O angles at the upper layer were observed. The $RuO_4$–(2×1) surface reconstruction is thermally stable.

Let us now consider the pseudocapacitive properties of studied $RuO_2$ reconstructions. Previous studies [28,29] concluded that surface redox reaction will not contribute to capacitance of cathode material, because the calculated voltage is above the oxygen evolution potential (OEP) for different numbers of adsorbed hydrogen atoms. [28,29] However, experimental work [23] reported that redox reaction should be responsible for pseudocapacitance. To resolve this, we calculated the voltages for $Ru_4O_9$–(1×1) and $RuO_4$–(2×1) reconstructions and prove that surface redox reaction takes place on the new $RuO_4$–(2×1) reconstruction. Here we considered OEP as a boundary value of voltage applied to the whole system. For ideal systems, where overpotentials are not considered, OEP is 1.23 V. If voltage, calculated by eq. (8), is less than 1.23 V, then one observes a predominant influence of hydrogen intercalation into the cathode surface, which would contribute to pseudocapacitance. If, on the other hand, the calculated voltage is > 1.23 V, then water splitting takes place and no proton adsorption or intercalation happen.

To calculate the voltage, we considered adsorption of hydrogen atoms on the $Ru_4O_9$–(1×1) and $RuO_4$–(2×1) surfaces. For the $Ru_4O_9$–(1×1) reconstruction, the most favorable positions of hydrogen atoms shown in Figure 5a were taken from Refs. [28,29,41]. The energies of hydrogen adsorption agree well with Ref. [41].All possible positions of hydrogen atoms (with the total number of atoms from 1 to 6) on the $RuO_4$–(2×1) surface were considered, and only the most favorable ones are shown in Figure 5b.

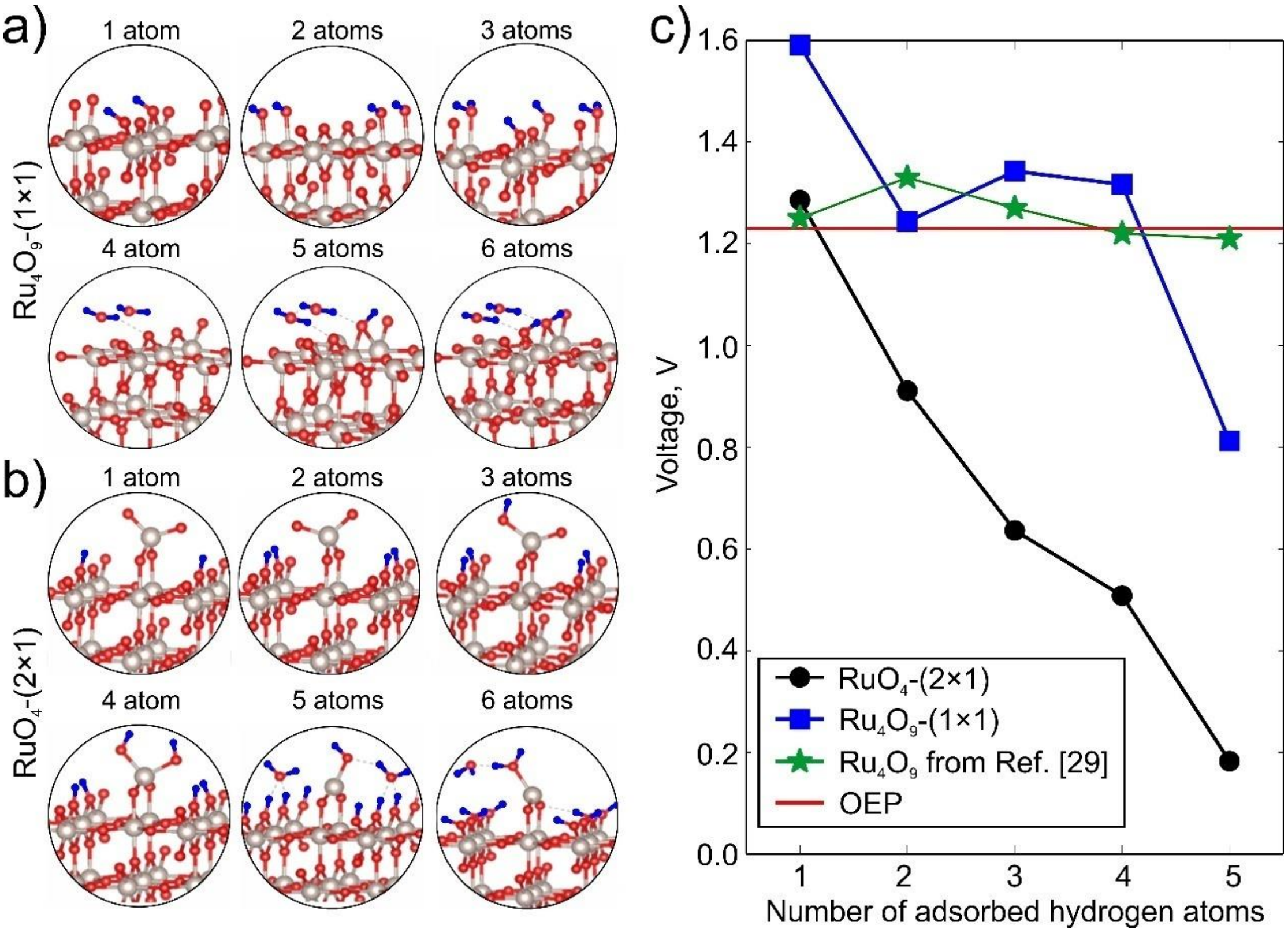

**Figure 5.** Configurations of adsorption sites of hydrogen atoms on the a) $Ru_4O_9$–(1×1) and b) $RuO_4$–(2×1) reconstructions with the total number of hydrogens atoms from 1 to 6. Ru atoms are grey, oxygen is red, and hydrogen is blue; c) Calculated voltage as a function of number of adsorbed hydrogen atoms for $RuO_4$–(2×1) (black color) and $Ru_4O_9$–(1×1) (blue color). Green stars are reference data from Ref. [29]. Oxygen evolution potential is shown by horizontal red line.

Using eqs. (8) and (9) we calculated the voltages of $RuO_4$–(2×1) reconstruction and $Ru_4O_9$–(1×1), compared to reference data from Ref. [29] (see Figure 5c). Calculated values of voltage for $Ru_4O_9$–(1×1) reconstruction are above or very close to OEP, which is in good agreement with Ref. [29] (green stars in the Figure 5c).

In stark contrast, $RuO_4$–(2×1) reconstruction with more than one adsorbed hydrogen atoms shows voltage below the OEP. This means that our new $RuO_4$–(2×1) reconstruction will adsorb hydrogen better than previously proposed [28] $Ru_4O_9$–(1×1) (see blue curve with squares).Such behavior explains and confirms the contribution of the surface redox reaction to the pseudocapacitance of $RuO_2$electrodes. We recall that $RuO_4$–(2×1) reconstruction is the dominant one at normal conditions (Figure 4).

In conclusion, we studied stable reconstructions of (110) surface of rutile $RuO_2$using global optimization algorithm USPEX. We found several new reconstructions, as well as all previously proposed ones. Predicted stable $RuO_4$–(2×1) reconstruction is found to be thermodynamically stable at normal conditions, and generally at oxygen-rich conditions. Simulated STM image of $RuO_4$–(2×1) reconstruction perfectly matches the experimental STM image. Calculated voltage for adsorption of hydrogen on the new $RuO_4$–(2×1) surface reconstruction is lower than oxygen evolution potential (OEP), and this result indicated the importance of the surface redox reaction to pseudocapacitance of $RuO_2$ cathodes.

**Methods**

Stable reconstructions of (110)-$RuO_2$ surface were predicted using first-principles evolutionary algorithm (EA) as implemented in the USPEX code, [30–33] where 4 different multiplications of unit cell were considered, namely (1×1), (1×2), (2×1) and (2×2). Here, evolutionary searches were combined with structure relaxations using density functional theory (DFT) [42,43] within the spin-polarized generalized gradient approximation (Perdew-Burke-Ernzerhof functional), [44] and the projector augmented wave method [45,46] as implemented in the VASP [47–49] package. The plane–wave energy cutoff of 500 eV, and *k*-mesh of $0.05 \times 2\pi/\text{Å}$ resolution ensure excellent convergence of total energies. During structure search, the first generation was produced randomly, and succeeding generations were obtained by applying 40% heredity, 10% softmutation, 20% transmutation operations, respectively and 30% using random symmetric algorithm. [50,51] Each of the considered supercells contained a vacuum layer of 15 Å and a substrate slab of 2 $RuO_2$ layers (6 Å) with atoms in the topmost 3 Å allowed to relax. We also performed additional calculations of slabs with thickness increased up to 12 Å, and only the bottom layer was kept fixed to obtain more accurate surface energies for stable (110)-$RuO_2$ reconstructions. No significant differences were found, which ensures reliability of our calculations.

For calculation of hydrogen adsorption on the predicted surface reconstructions, structure relaxation was carried out until the maximum net force on atoms became less than 0.01 eV/Å. The Monkhorst–Pack scheme [52] was used to sample the Brillouin zone, using 6×6×1 *k*-points mesh and the plane–wave energy cutoff was set to 500 eV.

For variable-composition search of optimal surface reconstructions, it is important to set boundary values of chemical potentials, which are related to the free energies of bulk Ru, $O_2$ molecule and bulk rutile-typeRuO$_2$. [53,54]

For the case of $RuO_2$, the surface energy can be written in the following manner:

$$G_s(T,P) = \frac{1}{N}[G^{slab}(T,P,N_{Ru},N_O) - N_{Ru}\mu_{Ru}(T,P) - N_O\mu_O(T,P)], \quad (3)$$

where $G_s(T,P)$ is surface energy per unit cell, $G^{slab}(T,P,N_{Ru},N_O)$ is the Gibbs free energy per cell of surface, which can be approximated as the total energy at 0 K, [20] $N = m{\times}n$ for an $m{\times}n$ surface supercell and serves as a normalization factor, $N_{Ru}$, $\mu_{Ru}$ and$N_O$, $\mu_O$ are the number and chemical potential of Ru and O atoms in the cell, respectively. In this approximation, temperature dependence is explicitly taken into account only for the chemical potential of oxygen (other values being much less dependent on the temperature).

Chemical potentials in equilibrium with of $RuO_2$ substrate are related through:

$$\mu_{Ru}(T,P) + 2\mu_O(T,P) = G_{RuO_2}(T,P), \quad (4)$$

where $G_{RuO_2}(T,P)$ is Gibbs free energy of bulk $RuO_2$. The surface energy can be recast in a form with only one variable chemical potential:

$$G_s(T,P) = \frac{1}{N}\left[G^{slab}(T,P,N_{Ru},N_O) - N_{Ru}G^{bulk}_{RuO_2}(T,P) - (N_O - 2N_{Ru})\mu_O(T,P)\right]. \quad (5)$$

Regarding physical bounds on chemical potentials, the chemical potential of Ru during crystallization on substrate was taken as lower limit and chemical potential when $O_2$ molecule goes away from the substrate was taken as upper limit. So the final relation, which defines the physically meaningful range of chemical potentials, has the following form:

$$\frac{1}{2}\Delta G_f(T,P) + \frac{1}{2}E_{O_2} \le \mu_O(T,P) \le \frac{1}{2}E_{O_2}, \quad (6)$$

where $E_{O_2}$ is total energy of oxygen molecule,$\Delta G_f(T,P)$ is the formation energy of bulk rutile-type $RuO_2$ from gas phase, equals 3.3 eV, which is in a good agreement with experimental value of 3.16 eV at 1000 K. [12,55] Above 1000 K the formation energy can reach the value of 3.2 eV.

Stability of different structures can be compared using equation (5) by plotting $G_s$ as a function of $\mu_O$ as shown in Figure 2a. Each structure corresponds to a line on the phase diagram. A complementary and equivalent way to determine stability is to plot the convex hull diagram (see Figure 2b), in Δ*E*-Δ*N* axes, [33] where

$$\Delta E = \frac{1}{N}\left[G^{slab}(T,P,N_{Ru},N_O) - N_{Ru}G^{bulk}_{RuO_2}(T,P)\right] \text{ and } \Delta N = \frac{1}{N}(N_O - 2N_{Ru}). \quad (7)$$

The calculation of electrode voltages was done using free energies of the surface with hydrogen adatoms on it. [29,41] The voltage can be calculated using other methods, i.e. joint density functional theory (JDFT), [56] which considers electrode-electrolyte interaction and overpotential influence. Another method considers *pH* and work function of surfaces. [24] However, all these methods strongly depend on the surface reconstruction. We calculate voltage of electrode, using method proposed by Liu et al., [29] which can determine the contribution of redox reaction to pseudocapacitance. The voltage was calculated by using the following equation:

$$V(n) = -\frac{\Delta G_H(n+1) - \Delta G_H(n)}{q_e}, \quad (8)$$

$$\Delta G_H(n) = G^{surf}_{RuO_2+nH} - G^{surf}_{RuO_2} - \frac{n}{2}G_{H_2}, \quad (9)$$

where *n* is the number of adsorbed or intercalated hydrogen atoms, $G^{surf}_{RuO_2+nH}$ is the free energy of surface with n adsorbed hydrogen atoms, $G^{surf}_{RuO_2}$ is the surface free energy, $G_{H_2}$is free the energy of $H_2$ molecule in a gas phase and *V*(*n*) is voltage as a function of the number of protons (hydrogen atoms) adsorbed on the surface or intercalated in the material. The voltage was calculated for the $RuO_4$–(2×1) and doubled cell of $Ru_4O_9$–(1×1) due to different sizes of considered unit cells. The hydrogen atoms (from 1 to 6 atoms) were adsorbed on different positions as was done in previous studies. [29,41] The calculated adsorption energies and voltages agree well with reference data. [29,41]

**Acknowledgements**

The work was supported by Russian Science Foundation (№ 16-13-10459). Calculations were performed on the Rurik supercomputer at MIPT. H.A.Z. would like to thank the Young Scientist Support Program (YSSP) headed by the patronage of the President of the Republic of Armenia, and Skoltech for supporting his stay at Skoltech.

**Author contributions statement**

H.A.Z. performed all the calculations presented in this article with help from A.G.K. Research was designed by A.R.O. H.A.Z., A.G.K. wrote the first draft of the paper and all authors contributed to revisions.

**Competing interests**

The authors declare no competing financial interests.